\documentclass[fleqn,usenatbib]{mnras}

\usepackage{newtxtext,newtxmath}
\usepackage[T1]{fontenc}
\usepackage{comment}

\DeclareRobustCommand{\VAN}[3]{#2}
\let\VANthebibliography\thebibliography
\def\thebibliography{\DeclareRobustCommand{\VAN}[3]{##3}\VANthebibliography}

\usepackage{graphicx}	% Including figure files
\usepackage{amsmath}	% Advanced maths commands
\usepackage{color}

\usepackage{xcolor}
\definecolor{purple2}{rgb}{0.5, 0.0, 0.5}

\newcommand{\mrm}[1]{\mathrm{#1}}
\newcommand{\nuc}[2]{$\mrm{^{#2}#1}$}

\title[$^{48}$V gamma-rays from SNe]{Prospects of direct detection of $^{48}$V gamma-rays from thermonuclear supernovae}

\author[F. H. Panther et al.]{
Fiona H. Panther$^{1,2,3}$\thanks{E-mail: fiona.panther@uwa.edu.au},
Ivo R. Seitenzahl$^{2}$,
Ashley J. Ruiter$^{2,3}$,
Thomas Siegert$^4$,
Stuart Sim$^5$,
Roland M. Crocker$^6$
\\
$^{1}$Department of Physics, University of Western Australia, Crawley WA 6009, Australia\\
$^{2}$School of Science, University of New South Wales, Australian Defence Force Academy, Canberra, ACT 2600, Australia\\
$^{3}$Australian Research Council Centre of Excellence for Gravitational Wave Discovery (OzGrav)\\
$^4$ Center for Astrophysics and Space Sciences, University of California, San Diego, 9500 Gilman Dr., La Jolla, CA 92093-0424, USA\\
$^5$Astrophysics Research Centre, School of Mathematics and Physics, Queen’s University Belfast, Belfast BT7 1NN, UK\\
$^6$The Research School of Astronomy and Astrophysics, Mount Stromlo Observatory, Australian National University, Canberra, ACT 2611, Australia.\\
}

\date{Accepted XXX. Received YYY; in original form ZZZ}

\pubyear{2021}

\begin{document}
\label{firstpage}
\pagerange{\pageref{firstpage}--\pageref{lastpage}}
\maketitle

% Abstract of the paper
\begin{abstract}
Detection of gamma-rays emitted by radioactive isotopes synthesized in stellar explosions can give important insights into the processes that power transients such as supernovae, as well as providing a detailed census of the abundance of different isotope species relevant to the chemical evolution of the Universe.
Observations of nearby supernovae have yielded observational proof that $^{57}$Co powered the late-time evolution of SN1987A's lightcurve, and conclusive evidence that $^{56}$Ni and its daughter nuclei power the light curves of Type Ia supernovae.
In this paper we describe the prospects for detecting nuclear decay lines associated with the decay of $^{48}$V, the daughter nucleus of $^{48}$Cr, which is expected to be synthesised in large quantities - $M_{\mathrm{Cr}}\sim1.9\times10^{-2}\,\mathrm{M_\odot}$ - in transients initiated by explosive helium burning ($\alpha$-capture) of a thick helium shell.
We calculate emergent gamma-ray line fluxes for a simulated explosion model of a thermonuclear explosion of carbon-oxygen white dwarf core of mass $0.45\,M_{\odot}$ surrounded by a thick helium layer of mass $0.21\,M_{\odot}$.
We present observational limits on the presence of $^{48}$V in nearby SNe Ia 2014J using the \textit{INTEGRAL} space telescope, excluding a $^{48}$Cr production on the surface of more than $0.1\,\mathrm{M_{\odot}}$.
We find that the future gamma-ray mission AMEGO will have an approximately 5 per cent chance of observing $^{48}$V gamma-rays from such events during the currently-planned operational lifetime, based on our birthrate predictions of faint thermonuclear transients.  
We describe the conditions for a $3\sigma$ detection by the gamma-ray telescopes \textit{INTEGRAL}/SPI, COSI and AMEGO. 
\end{abstract}

% Select between one and six entries from the list of approved keywords.
% Don't make up new ones.
\begin{keywords}
supernovae: general -- gamma-rays: general -- nucleosynthesis
\end{keywords}

%%%%%%%%%%%%%%%%%%%%%%%%%%%%%%%%%%%%%%%%%%%%%%%%%%

%%%%%%%%%%%%%%%%% BODY OF PAPER %%%%%%%%%%%%%%%%%%

\section{Introduction}
Detections of gamma-ray line emission at MeV energies allows us to directly observe nucleosynthesis, the process by which the chemical elements are created in stars and stellar end products.
Optical spectra - and to some extent photometry - allow us to understand broadly the chemical composition and  physical state of material produced by nucleosynthesis, including parameters like temperature and ionization state.
However, optical spectra do not permit the observer to know the isotopic composition of the material they are observing. 
Observations of gamma-ray spectra can provide direct information about the yield of different isotopes produced by steady-state or explosive nucleosynthesis processes.
Understanding the details of nucleosynthesis, and the various abundances of different isotopes that are produced in this process, is critical in enabling us to understand stellar evolution and chemical evolution over cosmic time \citep[see][for reviews]{DiehlHartmann, Kobayashi2020}.

Of particular interest is the process of explosive nucleosynthesis that occurs at the end of the lives of stars.
Small changes in the properties of the explosion progenitors, the conditions and realization of the explosion can yield significant differences in how explosive nucleosynthesis proceeds and in the final isotopic composition of the explosion ejecta.
Such model dependent differences in the yields of the synthesized radioisotopes have been proposed as discriminants between competing explosion models for Type Ia supernovae (SNe~Ia), see e.g.\ \citet{seitenzahl2013a} and \citet{seitenzahl2015} for the example of radioactive $^{55}$Fe or e.g.\ \citet{seitenzahl2009} and \citet{roepke2012} for the $^{57}$Ni/$^{56}$Ni ratio, see also \citet{Siebert2020}, \citet{Lach2020}.

Observing the prompt emission gamma-ray lines of radioactive material within the first few hours to hundreds of days post-explosion enables one to understand both the conditions of the explosion and the mechanism that powers other observables such as optical lightcurves.
For example, in `normal' SNe Ia \citep{Branch92}, the decay of radioisotope $^{56}$Ni to $^{56}$Co powers the lightcurve for the first 10 days after maximum light.
%
%After 
From 10 days post-maximum, the optical lightcurve is powered by the decay of $^{56}$Co to $^{56}$Fe. 
This mechanism was first proposed as the ultimate source of power radiated in the lightcurves of SNe Ia in the 1962 PhD thesis of Titus Pankey Jr \citep{PankeyPhD}, a theory that was subsequently independently discovered and expanded upon in \cite{Colgate1969}.
 The properties and evolution of these lightcurves forms the basis for our understanding of dark energy \citep{Phillips93}.

In 2014, the first direct observational evidence for $^{56}$Ni decay as the mechanism that powers SNe Ia was gathered, confirming Pankey and Colgate's theories.
Gamma-ray lines from $^{56}$Ni and $^{56}$Co decay were observed in the nearby Type Ia supernova SN2014J \citep{Churazov14, Diehl2014J} by the two main instruments, SPI and ISGRI, aboard the \textit{INTEGRAL} satellite \citep{INTEGRALmission}.
Detection of these gamma-ray lines also allowed for an accurate determination of the $^{56}$Ni mass produced in the explosion \citep{Churazov14, Diehl2014J,Isern2016}.
Previous to this measurement, gamma-ray derived limits on the production of $^{56}$Ni had been derived from SN2011fe, another nearby SNe Ia \citep{Isern2011, Isern2014}.
Even earlier, COMPTEL observations of SN~1991T have led to claims of detection of $^{56}$Co gamma-rays at low significance near the detection threshold \citep{morris1995, morris1997}, but these results are contested \citep{lichti1994, leising1995}.

%While normal SNe Ia are understood to be powered by $^{56}$Ni, synthesised during explosive carbon burning in the progenitor star(s), other types of supernovae may be powered by different radioisotopes.
%
In the nearby core-collapse supernova, SN1987A, at late times ($>800\,\mathrm{days}$) the light curve stayed bright longer than predicted under the assumption that the lightcurve was powered by the $^{56}$Ni decay chain alone \citep{Kurfess1992}.
Subsequent observations discovered gamma-ray lines produced in the decay of $^{57}$Co - the first detection of gamma-rays associated with nucleosynthesis beyond the Galaxy \citep{Kurfess1992}.
The mass of $^{57}$Co determined was sufficient to power the late time light curve \citep{Clayton1992}, and initial tension \citep{suntzeff1991,suntzeff1992} between the $^{57}$Co masses inferred from the direct detection results and the light-curve analysis were reconciled when the effects of freeze-out \citep{fransson1993,fransson2015} and internal conversion electrons \citep{seitenzahl2009, seitenzahl2014} are taken into account.

While normal SNe Ia were initially thought to be powered exclusively by the $^{56}$Ni chain, deep observations of nearby SNe Ia (SN 2003hv, SN 2011fe, SN 2012cg, SN 2013aa, SN2014J, ASASSN-14lp, SN2015F) revealed similar slow-downs of their light curves at late-times \citep{leloudas2009, graur2016, graur2018a, graur2018b, kerzendorf2017, dimitriadis2017, shappee2017, yang2018, jacobson-galan2018, li2019}, generally attributed to the dominant contribution of $^{57}$Co decay to the heating  at late times.

Such observations represent just the beginning of our ability to directly interrogate the process of nucleosynthesis in supernovae and other cosmic explosions.
Another isotope of particular interest is associated with the synthesis and decay of $^{48}$Cr.
The isotope $^{48}$Cr is synthesised during explosive $\alpha$-capture, and the short half-life of this radionuclide and its daughter $^{48}$V has been proposed as the source that powers the rapidly-declining lightcurves of SN2005E-like supernovae \citep{Waldman2011}, a subclass of SNe first described by \cite{Perets10} and subsequently studied in more detail by \cite{Kasliwal2012}, \cite{Yuan2013} and \cite{Lyman2013}. The synthesis of $^{48}$Cr in helium detonations and deflagrations has also been investigated by \cite{WK11}, who demonstrated that significant yields of $^{48}$Cr is expected in a variety of models - particularly helium deflagrations - while models that resemble more typical SNe Ia produce substantially less $^{48}$Cr. Thus, production of large quantities of $^{48}$Cr in thermonuclear supernovae is a characteristic of the presence of a thick helium layer in the progenitor.

The presence of $^{48}$Cr and $^{48}$V as the dominant isotopes that power the optical lightcurve of supernovae results in spectroscopic and photometric peculiarities - notably that they are sub-luminous and fast declining compared to `normal' SNe Ia. This has been noted in several works in which theoretical models of supernovae that produce significant quantities of these radioisotopes were investigated, including \cite{WK11, Waldman2011, Sim2012} The rapidly declining lightcurves of these models are particularly distinctive, following the radioactive decay of the short-lived $^{48}$Cr decay chain.
It has been proposed, furthermore, that there may be features associated with these isotopes present in the optical spectra of these events \citep{WK11}. However, it unfortunately appears that potential spectroscopic features that arise from 
%the presence of 
these atoms are difficult to discriminate from 
those due to
species with similar atomic numbers, namely Ca {\sc ii} and Ti {\sc ii}, at $\sim 4000\,$\AA. 
%
%Consequently, gamma-rays provide definitive evidence of the presence of $^{48}$Cr and $^{48}$V as the dominant source of radioactive heating in the ejecta, as we demonstrate in this work.
This serves as an extra motivation to consider, as we do here,
the idea that gamma-rays may provide definitive evidence of the presence of $^{48}$Cr and $^{48}$V as the dominant source of radioactive heating in the 
ejecta of transients whose progenitors have thick He shells.

$^{48}$V is of particular interest to observers of gamma-ray lines as it is not only indicative of explosive $\alpha$-capture in the progenitor explosion, but has an intermediate half-life of $T_{1/2}(\mathrm{^{48}V}) = 15.97\,\mathrm{d}$ that makes it a compelling observational target.
At present, the only telescope capable of observing gamma-ray lines such as those produced by the $^{56}$Ni decay chain, $^{57}$Co and $^{48}$V is \textit{INTEGRAL}/SPI.
\textit{INTEGRAL}/SPI \citep{SPI} is ideally suited to observing nuclear decay lines due to its high spectral resolution ($\sim2.7\,\mathrm{keV}$ at $\sim20\,\mathrm{keV}\,-\,8\,\mathrm{MeV}$ energies).
However, observations of supernovae in even nearby galaxies are limited by the sensitivity of the SPI instrument.
The Compton Spectrometer and Imager (COSI) \citep{Tomsick2019}, a Compton telescope, is currently under development.
COSI will boost the sensitivity to gamma-ray lines at $\sim\,\mathrm{MeV}$ over SPI thanks to improved background rejection.
The capabilities of compact Compton telescopes, such as COSI have been demonstrated through imaging and spectroscopic measurements of the Galactic $511\,\mathrm{keV}$ positron-electron annihilation line during a balloon flight of the instrument in 2016 \citep{Kierans2016, Kierans2020,Siegert2020}.
Other future gamma-ray observatories have also been proposed.
One such observatory is AMEGO, the All-Sky Medium Energy Observatory \citep{AMEGOmission}, a Compton telescope proposed for launch in the early 2030s.
AMEGO will combine high sensitivity between $200\,\mathrm{keV}\,-\,10\,\mathrm{MeV}$ ($3\sigma$ line sensitivity at $1\,\mathrm{MeV}$ of  $4\times10^{-6}\,\mathrm{ph\,s^{-1}\,cm^{-2}}$, an $\sim2$ order of magnitude improvement on \textit{INTEGRAL}/SPI) with a wide field of view with a spectral resolution of $\sim1\,\mathrm{per\,cent}$ at $1\,\mathrm{MeV}$.

In this paper we discuss the prospects for detecting prompt gamma-ray emission associated with nucleosynthesis of radioactive $^{48}$V in subluminous thermonuclear events based on the hydrodynamic and nucleosynthesis modelling of such explosions in \cite{Sim2012}.
We discuss the rates and implications of such an event for current and future telescopes capable of detecting MeV gamma-ray line emission, including \textit{INTEGRAL}/SPI, COSI and AMEGO.
\section{Theory}
\subsection{Nucleosynthesis}
\label{sec:Vprod} % used for referring to this section from elsewhere
Explosive nuclear burning in detonations in helium layers on top of WDs is typically incomplete, in the sense that rapid expansion of the hot ashes leads to a freeze-out of the nuclear fusion reactions before nuclear statistical equilibrium is achieved \citep[e.g.][]{townsley2012,moore2013}.
Consequently, compared to explosive carbon-oxygen burning \citep[e.g.][]{livne1995,seitenzahl2013b}, the ashes of helium-detonations are over-abundant in $\alpha$-isotopes such as $^{36}$Ar, $^{40}$Ca, $^{44}$Ti, $^{48}$Cr, or $^{52}$Fe \citep[e.g.][]{woosley1994,seitenzahl2017}. The most abundant of these $\alpha$-isotopes is determined by the conditions of the explosion, notably the pre-explosion density (which in turn determines the peak temperature), with successive $\alpha$-captures proceeding to the heaviest nucleus that can be formed on hydrodynamic timescales -- in this case, $^{48}$Cr. Pollution of the helium layer by admixture of e.g. carbon or oxygen can also result in an ``alpha-limited burn" and great enhancement of ``stagnation nuclei", with a ratio of $^{4}$He to $^{12}$C around 4:1 by mass resulting in the greatest production of $^{48}$Cr \citep[see section 4.3 of][]{gronow2020}.
These scenarios are unique to helium detonations. Another mode by which these $\alpha$-isotopes can be synthesised is the so-called $\alpha$-rich freeze out, which occurs during the carbon detonation or deflagration processes. However, typical masses of $\alpha$-isotopes produced in this process are much lower, with final yields of ${\sim}10^{-6}\,\mathrm{M}_{\odot}$ of $^{48}$Cr in deflagration \citep[e.g.][]{fink2014} and ${\sim}10^{-4}\,\mathrm{M}_{\odot}$ of $^{48}$Cr in delayed-detonation \citep[e.g.][]{seitenzahl2013b} models, and these explosions are characterized as normal SNe Ia based on their lightcurves as the dominant nuclear product is $^{56}$Ni. The helium detonation with which we concern ourselves here is a rarer event, with both spectroscopic and photometric peculiarities described in \cite{Sim2012}, where our model is first presented.

 $^{48}$Cr itself is a poor candidate for direct detection with gamma-ray telescopes, since the short half-life of ${\sim}24\,\mathrm{h}$ means that most decays will occur at very high optical depth. 
 Further, any gamma-ray observation would have to be triggered by an initial optical detection of the supernova.
By the time the hypothetical supernova becomes optically bright, 2-3 days post-explosion, the majority of the synthesised $^{48}$Cr will already have decayed to $^{48}$V.

The decay product $^{48}$V is a radioisotope of vanadium (atomic number 23), on the proton-rich side of the valley of stability, with a half-life of 15.9735(25) days \citep{burrows2006}.
%has a half-life sufficiently long that gamma-rays emitted from its decay \rmc{might} be detected \rmc{in principle} for around one month post-explosion.
%
With a ground state to ground state Q-value of 4012.3(24) keV, the decay to stable $^{48}$Ti occurs roughly at 50\% via electron capture and 50\% via $\beta^+$-decay. 
The decay radiation includes a number of positrons, electrons (both Auger and internal conversion), X-rays, and gamma-rays (both from nuclear transition as well as the 511\,keV annihilation line).
A very limited selection of the most prominent decay products and their intensities is summarized in table~\ref{tab:table2}. In this paper, we will focus on the emission of the prominent gamma-rays at $983\,\mathrm{keV}$ and $1312\,\mathrm{keV}$.

The longer half-life of $^{48}$V, coupled with its production at relatively low-optical depth in the helium-rich surface layers of sub-luminous thermonuclear supernovae \citep[e.g.][]{Waldman2011,Sim2012}, means that these two main nuclear gamma-rays are promising direct-detection prospects for future planned missions (see section \ref{subsec:det}).

\begin{table}
    \centering
    \begin{tabular}{ccc}
    \hline
     Energy [keV] & Intensity [\%] & radiation type  \\
    \hline
    4.0 & 35.1 & e$^-$ (Auger K) \\
    4.505 & 2.9 & X-ray (Ka2) \\
    4.511 & 5.8 & X-ray (Ka1) \\
    290.3(11) & 49.9(4)& e$^+$ \\
    511.0 & 99.8 (8) & $\gamma$ (annihil.) \\
    983.525(4) & 99.98(4)& $\gamma$ \\
   1312.106(8) & 98.2(3) & $\gamma$ \\
    \hline
    \end{tabular}
    \caption{Select decay radiation of $^{48}$V. Source: \citet{burrows2006}}
    \label{tab:table1}
\end{table}

\subsection{Description of radiative transfer model}
The gamma-ray spectra used to investigate the detectability of $^{48}$V are obtained from the output of explosion simulations, nucleosynthesis and radiation transport simulations described by \citet{Sim2012}.

Our general model is based on the explosion simulations studied in \citet{Sim2012}. That work explored the outcome of detonations in isolated, low-mass carbon-oxygen white dwarfs (CO WDs) that are surrounded by massive  helium shells, regardless of how such a configuration may arise in nature. To assess how such a star may end up with such a configuration -- and quantify the birthrates of such systems, which we would expect to only arise in interacting binaries -- we use the {\sc StarTrack} binary evolution population synthesis code (see below). For the present work, we assume that CO WDs may accumulate a non-negligible amount of helium through rapid mass transfer, e.g. during a merger event with a helium-rich star (either a helium or HeCO `hybrid' white dwarf). We describe our method in Section 4. 

We utilize the edge-lit core detonation (ELDD) model of \cite{Sim2012}.
Specifically, we choose the ELDD-L model as it provides a benchmark for the emission of gamma-rays from the decay of $^{48}$V for supernovae that arise from helium detonations. 
In this model, the detonation of the helium layer ignites the CO core of the progenitor.
The parameters of the pre-explosion model can be found in Table 1 of \cite{Sim2012} and is replicated here in Table \ref{tab:table2} of this paper for clarity.

\begin{table}
    \centering
    \begin{tabular}{cc}
    \hline
    Parameter/Unit & Value\\
    \hline
    CO core central temperature & $5\times10^7\,K$\\
    CO core central density & $3.81\times10^6\,\mathrm{g\,cm^{-3}}$\\
    He layer base temperature & $2\times10^8\,K$\\
    He layer base density  & $0.592\times10^6\,\mathrm{g\,cm^{-3}}$\\
    CO mass & $0.45\,M_{\odot}$\\
    He layer mass & $0.21\,M_{\odot}$\\
    Total mass & $0.66\,M_{\odot}$\\
    $^{48}$V mass synthesised & $1.9 \times10^{-2}\,\mathrm{M_\odot}$\\
    \hline
    \end{tabular}
    \caption{Properties of the progenitor system of the studied explosion model. The masses are not independent but are determined from the temperature and density parameters, and the equations of hydrostatic equilibrium. Source: \citet{Sim2012}}
    \label{tab:table2}
\end{table}

Details of the explosion simulations, nucleosynthesis and radiation transport are given by %both 
\cite{Sim2012} and \cite{Fink2010}.
Synthetic observables, including gamma-ray spectra, are derived from the simulations of the ELDD-L model %described in Sim+2012 
using the \texttt{ARTIS} radiation transport program \citep{Sim2007, Kromer2009}. For our analysis, we use the angle-averaged $\gamma$-ray synthetic spectrum for simplicity (note that the underlying explosion models are two-dimensional an therefore predict a degree of dependence on the observer orientation -- however, for such variations are modest and not the a dominant source of uncertainty for our estimates here).
In this model, the dominant radioactive species in the ejecta is $^{48}$Cr (and hence its daughter nucleus $^{48}$V) which is sythesised by the detonation of the He-layer ($M = 1.9\times 10^{-2}\,\mathrm{M_\odot}$).
Only a small mass of $^{56}$Ni is synthesised in the He-layer ($M = 7.6\times10^{-3}\,\mathrm{M_\odot}$) and a negligible quantity ($M = 1.5\times10^{-7}\,\mathrm{M_\odot}$) in the CO core.

At very early times ($\sim\,\mathrm{hrs}-\mathrm{day}$), there is a small contribution to the lightcurve from $^{52}$Fe, however the bolometric lightcurve is mainly powered by the decay of $^{48}$Cr and $^{48}$V in the early phase, between 3-30 days post-explosion .
Almost all radioactive material is located in the He-layer ejecta at low optical depths, and consequently some of the gamma-rays from these decays are able to escape.
The escaping gamma-rays give rise to line emission at $\sim\,\mathrm{MeV}$ energies (Table \ref{tab:table1}) that %can 
might be detected by space-based telescopes.

\section{Simulated Observations}
\begin{figure*}
\includegraphics[width=\textwidth]{{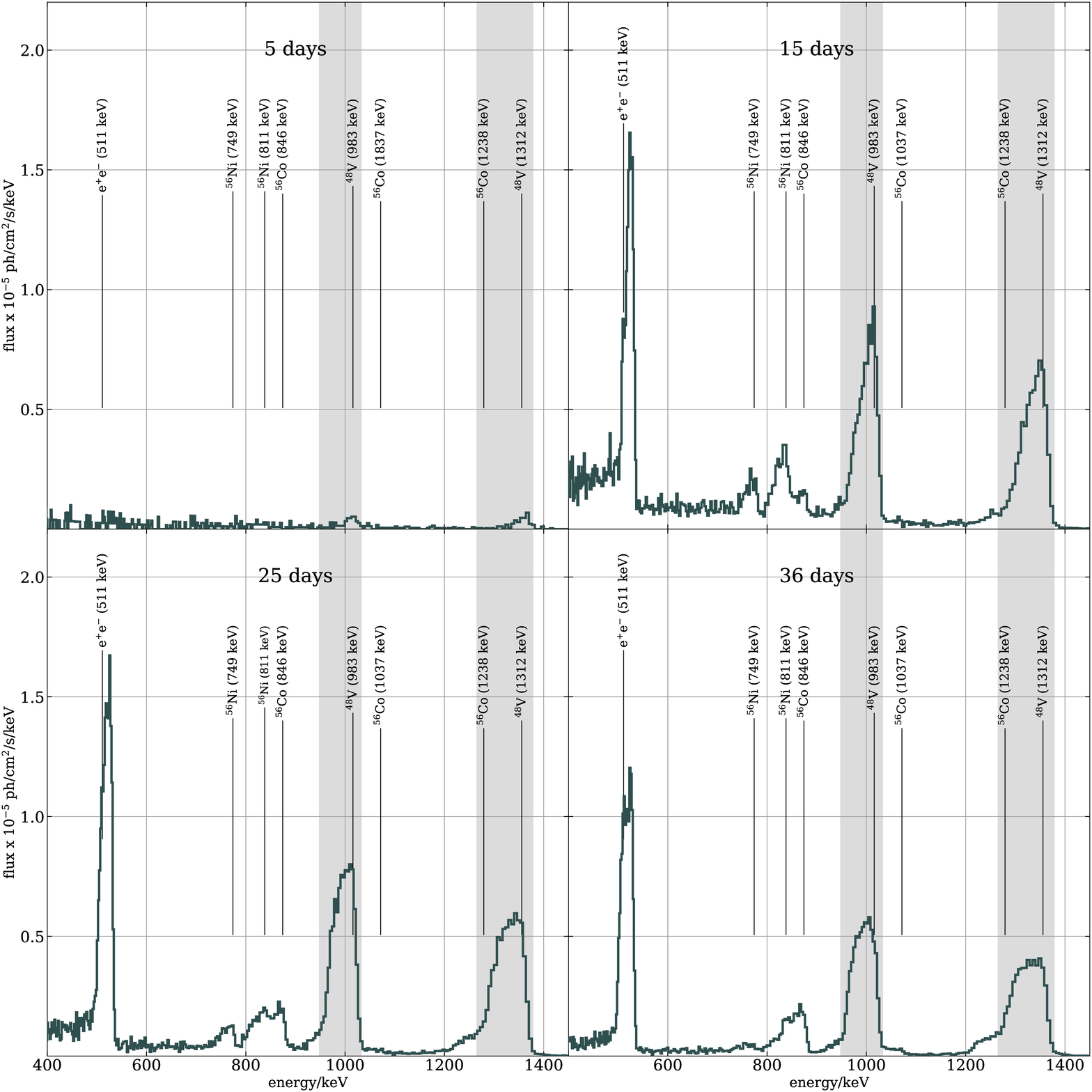}}
\caption{\label{fig:spec} Evolution of the gamma-ray spectrum of ELDD-L model at $\sim\mathrm{MeV}$ energies. The gamma-ray spectrum is shown at 5, 15, 25 and 35 days post-explosion. The shaded bands indicate the energies over which the line emission due to the decay of $^{48}$V is integrated to compute the lightcurve for each line. The line center is approximated to be blueshifted by 10,000 km/s with respect to the rest frame energy of each $^{48}$V gamma-ray line. The minimum of each energy band is then at -20,000km/s with respect to the line center, and the maximum energy is at +5000 km/s with respect to the line center. Other characteristic gamma-ray lines are indicated with their rest-frame energies included in brackets. Line emission from positron annihilation ($\sim 511\,\mathrm{keV}$) and decay of $^{56}$Ni and $^{56}$Co are also visible at later times.}
\end{figure*}

\begin{figure}
\includegraphics[width=\columnwidth]{{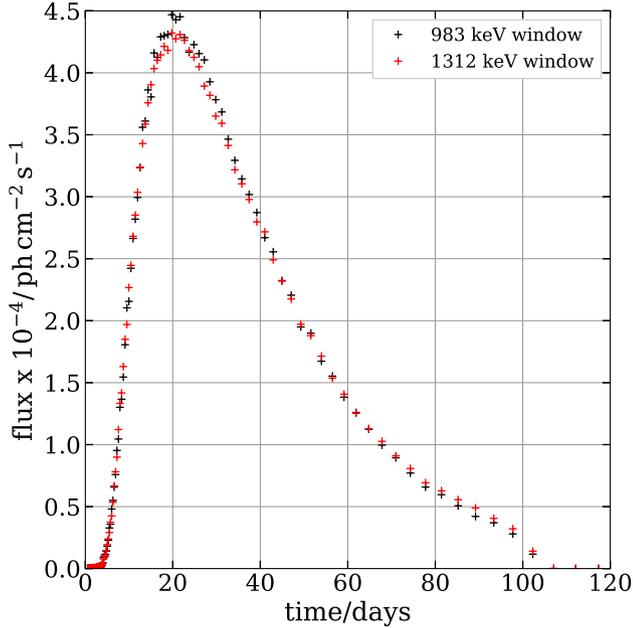}}
\caption{\label{fig:lightcurve}Gamma-ray lightcurve for ELDD-L model 983 keV and 1312 keV gamma-ray line emission in the energy bands defined in Fig \ref{fig:spec} for a supernova at a canonical distance of 1 Mpc. At 70 days post-explosion, $^{56}$Co emission becomes comparable to the strength of the $^{48}$V lines, resulting in the slight plateau visible in the lightcurve.
}
\end{figure}

\begin{figure}
\includegraphics[width=\columnwidth]{{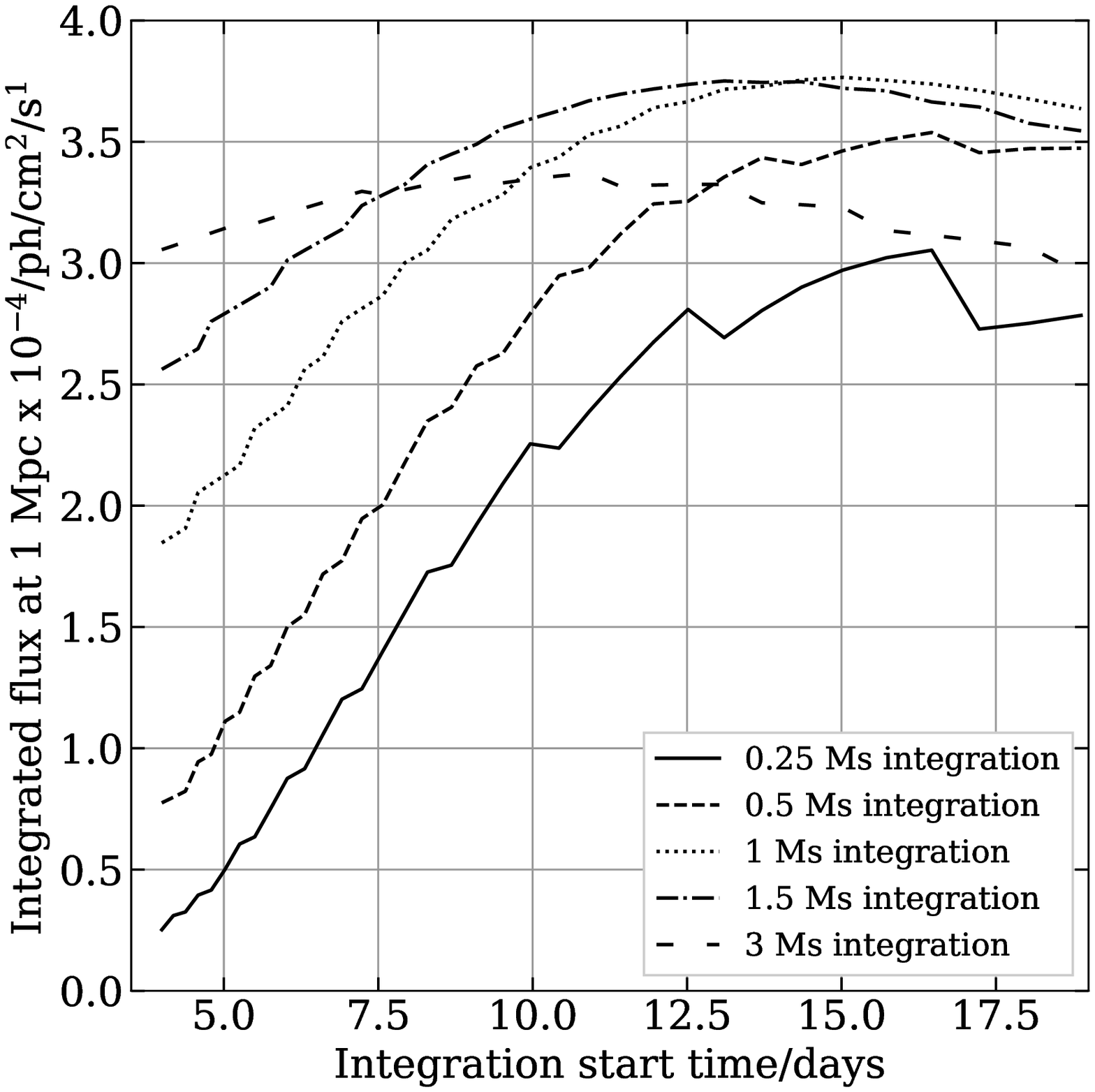}}
\caption{\label{fig:inttime}How the time-integrated flux of the $^{48}$V 1312 keV line for the ELDD-L model at 1 Mpc varies as a function of the time at which the integration is started. Observations that begin close to maximum light yield the maximum integrated flux for a given observation time. We also find that the optimal integration time is $\sim 1-1.5\,\mathrm{Ms}$.
}
\end{figure}

We simulate the gamma-ray emission from a supernova that results from the ELDD-L model of \cite{Sim2012} with the event located at a canonical distance of $1\,\mathrm{Mpc}$. The optical properties of this explosion model - spectra and lightcurve - are presented in detail in \cite{Sim2012}.
The spectra produced by the model are shown in Fig. \ref{fig:spec} at a range of different times post-explosion.
The two prominent emission lines result from the decay of $^{48}$V.
%
% The higher the gamma-ray energy, the lower the optical depth.
% %
% Hence the gamma-ray lines are observed to 'lean' toward higher energies.
%
Unlike normal SNe Ia, where $^{56}$Co dominates, emission from $^{56}$Co is very weak in comparison to the $^{48}$V line.

We use these spectra to compute the lightcurve of each emission line associated with $^{48}$V decay. 
The emission lines are numerically integrated in the window shown by the shaded region in Fig. \ref{fig:spec}.
These regions are defined as follows:
The offset of the line center with respect to the lab energy of each line, which represents the bulk motion of the ejecta, is blueshifted.
We determine this blueshift to be well-approximated at a velocity of $10,000\,\mathrm{km\,s^{-1}}$.
We then define the lower bound of the integration region to be redshifted by $20,000\,\mathrm{km\,s^{-1}}$ with respect to the line center for each emission line.
This redshift is due to a combination of the velocity of the ash and Compton scattering. 
The upper bound of the integration region is then defined to be blueshifted by $5,000\,\mathrm{km\,s^{-1}}$ with respect to the line center.
Defining the integration region in velocity space ensures both lines, which arise from the same material with the same kinematics, are treated identically. 
The bounds of the energy bands are found to be $\sim 950 - 1030\,\mathrm{keV}$ for the $983\,\mathrm{keV}$ window and $\sim 1265 - 1380\,\mathrm{keV}$ for the $1312\,\mathrm{keV}$ window.
We numerically integrate the flux in each of these windows at each time sample from 0 - 120 days via numerical integration, implemented using the Python \texttt{numpy} library function \texttt{numpy.trapz} where each bin is defined by the original binning of the spectrum. 
The effect of choosing these broad energy bands on the detection sensitivity is discussed in section \ref{subsec:det}. 

Using the gamma-ray lightcurve, we can determine how to achieve an observation that maximises the observable flux in a given observation time.
We consider observation times of $0.25, 0.5, 1, 1.5$ and $3\,\mathrm{Ms}$, typical of the time that would awarded to a target of opportunity observation.
We calculate the observed flux of the ELDD-L model at the canonical distance of 1 Mpc for each of these windows beginning at a different initial time $t_\mathrm{start}$ based on the lightcurve in Fig. \ref{fig:lightcurve}.
In Fig. \ref{fig:inttime} we demonstrate that for a given integration time, beginning the integration several days before the peak of the lightcurve (12-14 days) maximises the integrated flux that can be observed for a given integration time.
We also find that the maximum observable flux saturates with an integration time of $1-1.5\,\mathrm{Ms}$.
The subsequent calculations for the detectability of $^{48}$V are performed for an integration time of $1.5\,\mathrm{Ms}$ beginning at 12.5 days post-explosion.

\section{Prospects for detectability} \label{subsec:det}
The sensitivity of a gamma-ray instrument is dependent on several factors, including the integration time of the observation, the energy resolution of the instrument, the detector collecting area, and the background. 
In this work we will account for the impact of both the integration time of the observation and the energy resolution of the given instrument. 
We utilize published three-sigma sensitivity curves and detector resolution at $1\,\mathrm{MeV}$ for the narrow-line sensitivity of \textit{INTEGRAL}/SPI \citep[e.g.,][see also\footnote{\url{https://www.cosmos.esa.int/web/integral/observation-time-estimator}}]{Attie2003_SPI,Diehl2018_SPI} and AMEGO \citep{AMEGOmission}. The sensitivity of the COSI instrument was communicated via private communication with T. Siegert and the COSI collaboration.
For a given three-sigma, 1 Ms sensitivity, the three-sigma sensitivity $S_\mathrm{3\sigma}$ for an observation with a given time $t_\mathrm{o}$ Ms over a line with FWHM of $E_\mathrm{band}$ keV for an instrument with a narrow line sensitivity of $E_\mathrm{line}$ keV is given by
$$
S_\mathrm{3\sigma}(t_\mathrm{o}) =  S_\mathrm{3\sigma} \sqrt{\frac{1\,\mathrm{Ms} \times E_\mathrm{band}}{t_\mathrm{o}E_\mathrm{line}}}
$$
Because the observed emission line is significantly broader than the resolution of the telescope, there is a reduction in the sensitivity given by $f_\mathrm{broad} = \sqrt{\mathrm{FWHM}/E_\mathrm{line}}$ included in the determination of the 3-sigma sensitivity calculation.

We calculate the maximum distance $d_\mathrm{max}$ at which a three-sigma detection of each $^{48}$V line with an integrated flux of $F\,\mathrm{ph\,cm^{-2}\,s^{-1}}$ at the canonical distance of $1\,\mathrm{Mpc}$ can be made for a given exposure time (Table \ref{tab:table3}), where
$$
d_\mathrm{max}(t_\mathrm{o}) = 1\,\mathrm{Mpc}\times\sqrt{\frac{F}{S_\mathrm{3\sigma}(t_\mathrm{o})}}
$$

\begin{table*}
    \centering
    \begin{tabular}{cccc}
    \hline
    Exposure time (Ms) & \textit{INTEGRAL}/SPI $d_\mathrm{max}$ (Mpc) & COSI $d_\mathrm{max}$ (Mpc) & AMEGO $d_\mathrm{max}$ (Mpc)\\
    \hline
     & At $983\,\mathrm{keV}$ & \\
    \hline
    1 & 2.11 & 3.69 & 6.29 \\
    1.5 & 2.33 & 4.08 & 6.96\\
    \hline
     & At $1312\,\mathrm{keV}$ & \\
    \hline
    1 & 2.08 & 3.65 & 6.22 \\
    1.5 & 2.08 & 4.03 & 6.89 \\
    \end{tabular}
    \caption{Maximum distances (in Mpc) at which an observation of $^{48}$V decay from our model can be made, based on existing (for \textit{INTEGRAL}/SPI) and predicted (for COSI and AMEGO) narrow-line sensitivity of the instruments.}
    \label{tab:table3}
\end{table*}

To determine the rate of events that would yield a detection of $^{48}$V as a function of $d_\mathrm{max}$, we calculate the mass currently contained in stars in the local universe. A summary of measurements of the local stellar mass density is given in \cite{Karachentsev2018}. We utilize the MK11 measurement of \cite{Makarov2011} shown in Fig. 4 of \cite{Karachentsev2018} to compute the enclosed mass of stars as a function of distance.

As stated above, the binary population synthesis code \texttt{StarTrack} \citep{startrack2002, startrack2008, Ruiter2014} is used to estimate the number of star systems that could give rise to ELDD-L-like explosions. 
The primary goal of the \citet{Sim2012} study was to examine what a low-mass exploding CO WD may look like, and though CO WD masses were chosen based on population synthesis models, the final 1D hydrostatic configuration was not based on any specific binary evolution calculation. Our goal here however is to assess the likelihood of observing gamma ray emission from exploding white dwarfs of relatively low-mass, i.e., CO WDs that are too low-density to explode as `normal' SNe Ia \citep[see e.g.][]{Ruiter2020}. In order to achieve a situation where a relatively low-mass CO WD has a massive helium layer on its surface, the CO WD should be accreting from (or has recently accreted from) a star containing a substantial amount of helium on the surface. 
While certain binary configurations involving stable (Roche-lobe overflow on a nuclear or thermal timescale) mass transfer will lead to helium-burning on the WD surface (e.g. converting the helium to carbon, thereby removing helium), there are certain regions of parameter space during stable Roche-lobe overflow where helium will be accumulated on the CO WD and will not burn \citep{Piersanti2014}. Is is also plausible that a somewhat low-mass carbon-oxygen white dwarf with a rather thick helium surface layer could be realised through a merging event \citep{Crocker17}, though detailed studies of such a situation remain largely unexplored, particularly in 3D \citep[but see][]{pakmor2021}.

Our population of thermonuclear events should consist of both 1) CO WDs of low-enough mass such that they would produce very little radioactive nickel in a detonation and 2) a non-negligible layer of helium on the WD surface to allow for a shell detonation.  
We find that rapid mass transfer (e.g. a merger) between a carbon-oxygen white dwarf star and a helium-rich degenerate companion is the most likely scenario in terms of absolute numbers, given these two main constraints. This merger configuration, unlike the slow Roche-lobe overflow configuration where mass transfer proceeds on non-dynamical timescales, could plausibly lead to the formation of a star with an inner carbon-oxygen core of relatively low mass ($\lesssim 0.55\,\mathrm{M_\odot}$) surrounded by a He-rich outer layer. Systems with similar properties were already discussed in \citet{Crocker17} as plausible candidates for low-luminosity thermonuclear supernovae. 

Binary star systems leading to the formation of a CO WD that accretes helium-rich matter via slow Roche-lobe overflow from a white dwarf containing helium on its surface have been discussed in \citep{Ruiter2011,Ruiter2014}. While in those works it was assumed that mass transfer would be stable in most cases when the stellar mass ratio was above a certain threshold, e.g. when the primary WD is notably more massive than the secondary \citep{startrack2008}, it is possible that in such systems, mass transfer becomes unstable, and the two stars actually merge \citep[e.g. see][]{Shen2015}. However, the stable Roche-lobe overflow in the population synthesis studies discussed above found that the WD mass was typically larger than required for our study \citep[cf. fig 2][]{Ruiter2014}, thus we do not include these types of systems in our rate estimate.
In our estimate of birthrates of thermonuclear transients involving helium shell detonations, we consider mergers between CO WDs and WDs that are either `fully' helium-rich or contain helium-rich mantles on top of a CO core; HeCO `hybrid' WDs \citep{Iben1985}. 
It turns out that most merging systems consist of a CO WD (we set the upper mass limit to $0.55$ M$_{\odot}$ as discussed above) and a hybrid WD.  Such WDs are formed only in binary systems when a red giant branch star experiences substantial mass loss, in this case during a common envelope event. The most prominent formation channel leading to our desired binary configuration consists of a stable phase of Roche-lobe overflow (from the primary star in the Hertzsprung Gap and/or on the Red Giant Branch donating mass to the MS secondary) $\rightarrow$ a phase of unstable mass transfer (the now Red Giant secondary loses its envelope while the primary is a CO WD; the CE leaves behind a hydrogen-stripped, helium-burning star) $\rightarrow$ a final phase of stable Roche-lobe overflow but this time, the stripped He-burning star donates mass to the CO WD. At some stage the secondary evolves into a hybrid WD, and eventually the larger star (the hybrid) fills its Roche lobe. This time mass transfer is unstable, and the stars are presumed to merge. To translate our simulated events into predicted events, we calibrate our raw model data (${\sim} 3000$ systems) by tracking the mass born in stars for our simulation and scale this to the estimated mass in stars in the observed sample volume assuming a binary fraction of 70\%. We emphasize that our rate estimate should be taken as a lower limit, since we are not including the (more numerous) systems which are presumed to encounter stable Roche-lobe overflow in the code. 

%%%possibly needed in paper, but maybe not - This is more FYI: From a simulation of 12.8 million binaries, we find 2942 such systems with CO WD (with mass below $0.55$ M$_{\odot}$ at merger) merges with another WD that contains helium.

\begin{figure}
	\includegraphics[width=\columnwidth]{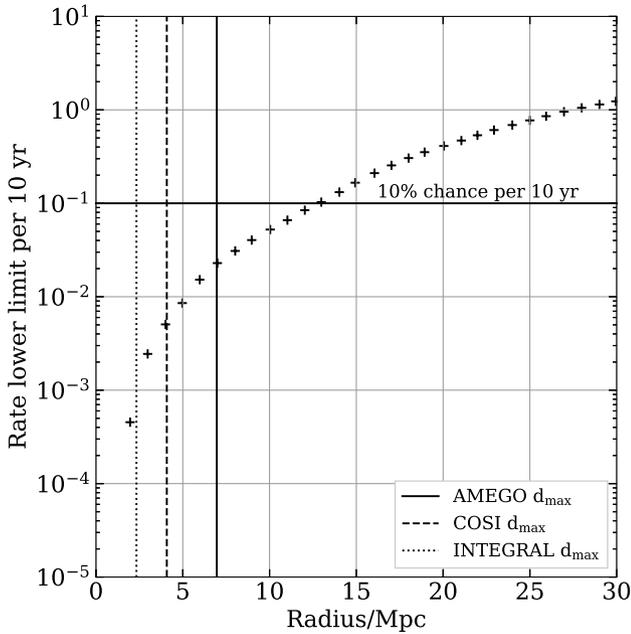}
    \caption{Rate per $10\,\mathrm{yr}$ of $^{48}$V-line emitting supernovae determined from binary population synthesis calculations and the stellar density in the local volume. Over a $10\,\mathrm{yr}$ mission we estimate a $\sim 5\,\mathrm{per\,cent}$ chance of observing such an event. While the probability of such an event is low, a nearby SNe Ia would provide interesting constraints on nucleosynthesis of $^{48}$V.}
    \label{fig:probrate}
\end{figure}

In Fig \ref{fig:probrate} we show the rate of such systems per ten years (an optimistic extended lifetime for space-based observatories) as a function of $d_\mathrm{max}$. 
These rates are a lower limit as we calculate rates based on the present-day mass in stars.
For the most optimal observation (1.5 Ms with AMEGO) we find a lower limit to the event rate to be ~0.02 per ten years (i.e. a 2 percent chance of a successful observation within a ten year mission lifetime.

\section{Observational limits from previous observations of SN2014J}
During the lifetime of \textit{INTEGRAL}, only two thermonuclear supernovae were close enough to be within reach for meaningful measurements: SN2011fe in M101 \citep{Nugent2011_SN2011fe} and SN2014J in M82 \citep{Fossey2014_SN2014J}. While SN2011fe has not been detected in gamma-rays \citep{Isern2011}, SN2014J merited an observation campaign for about half a year and led to the first detection of gamma-rays from \nuc{Co}{56} from the core of an exploding WD \citep{Churazov14,Diehl2015_SN2014J}. Early observations, only 16 days after the explosion date, also found gamma-rays from the decay of the shorter-lived parent-nucleus \nuc{Ni}{56}, which was interpreted as the production of $\sim 0.06\,\mrm{M_{\odot}}$ of \nuc{Ni}{56} on the surface of the WD \citep{Diehl2014J,Isern2016}. These early observations are also close to the optimal point for the search of \nuc{V}{48} (see Fig.\,\ref{fig:inttime}), so that we can compare our model expectations with this observation.

Between days 16.3 and 30 (i.e. $\sim 1.2\,\mrm{Ms}$), the 983\,keV and 1312\,keV lines are not detected. Assuming a line broadening of $10000\,\mrm{km\,s^{-1}}$, the $3\sigma$ flux limits during this period are $2.4 \times 10^{-4}\,\mrm{ph\,cm^{-2}\,s^{-1}}$ and $2.1 \times 10^{-4}\,\mrm{ph\,cm^{-2}\,s^{-1}}$ for the 983 and 1312\,keV line, respectively. Given the distance to SN2014J of $3.3\,\mrm{Mpc}$ \citep{Foley2014_SN2014J}, these limits are about 5--6 times larger than what would be expected in our model. This excludes a \nuc{Cr}{48} production on the surface of more than $0.1\,\mrm{M_{\odot}}$. While these values are hardly constraining, it is important to show the consistency between the expectations of this high-\nuc{Cr}{48}-yield model and observations, especially because SN2014J also showed significant contributions of nucleosynthesis products on its surface. Later observations, for example between day 50 and 70 after the explosion, are less constraining, even though the flux limits drop to $1.6 \times 10^{-4}\,\mrm{ph\,cm^{-2}\,s^{-1}}$ and $1.3 \times 10^{-4}\,\mrm{ph\,cm^{-2}\,s^{-1}}$ for the 983 and 1312\,keV line, respectively. 

We note that while \textit{INTEGRAL} is unable to provide measurements of emission from $^{48}$V decay in SN2014J (and hence constraints on the model of \cite{Diehl2015_SN2014J} which proposes the presence of helium-rich material in the explosion), had AMEGO or COSI been operational when the event occurred then this model could have been either confirmed or falsified, based on the above constraints from \textit{INTEGRAL}.

\section{Discussion}
The scope of this work is to investigate the feasibility of an observation of radioisotope decay that is characteristic of the thermonuclear explosion of massive helium shells. While the presence of $^{48}$Cr and its daughter nuclei have significant implications for the optical lightcurves and spectra of these supernovae \citep{WK11, Waldman2011, Sim2012}, definitive evidence for their presence as the dominant product of helium detonation (or deflagration) can be obtained through gamma-ray line observations. We note some potential avenues for detector improvement to enable such a detection, and the limitations of our rate estimations used in this work.

Increasing the sensitivity of detectors can have a significant impact on the probability of detecting gamma-ray emission from supernovae. For example, a factor of two improvement in the sensitivity of AMEGO at $1\,\mathrm{MeV}$ increases $d_\mathrm{max}$ to $\sim9\,\mathrm{Mpc}$. This would double the rate of observable events per ten years (see fig. \ref{fig:probrate}).

In this paper we consider gamma-ray observatories that can make both wide field-of-view observations of diffuse emission and observations of point sources such as supernovae. A factor of $\sim 6$ improvement of the sensitivity of an instrument to observe large areas at once, like AMEGO or COSI, and improved background suppression and rejection could enable serendipitous detection of gamma-rays from supernovae as far out as the Virgo cluster, at a distance of ${\sim}16.5\,\mathrm{Mpc}$.

AMEGO and COSI are optimized to observe large areas of the sky. Telescopes that are optimized to observe point sources of MeV gamma-rays are more ideally suited to carry out the observations described in this paper. 
For example, the proposed Lunar Occultation Observatory (LOX, \citealt{LOX}) uses lunar occultation as a method of observing gamma-rays from point sources. Lunar occultation enables detector background, which limits detector sensitivity, to be removed with much greater efficacy.
Coupled with the intrinsically lower background due to the proposed observatory's lunar orbit, LOX's sensitivity could enable the detection of $^{56}$Ni and $^{56}$Co in SNe Ia out to $\sim10\,\mathrm{Mpc}$.
Comparing the flux of the $^{56}$Ni and $^{56}$Co lines found in SN2014J in \cite{Diehl2014J} to that of the $^{48}$V emission in our analysis, we find that an observatory like LOX can detect $^{48}$V emission at comparable distances.

Alternative MeV gamma-ray telescopes optimized for observing point sources include Laue lenses, which use the principle of Laue diffraction of gamma-rays in a lens made of high-purity Ge crystals to focus gamma-rays into the detector plane. 
Laue lenses have been proposed for astronomy applications, particularly observing $\sim \mathrm{MeV}$ emission from point sources, however most designs for such a telescope are currently highly experimental.

The scope of this paper is to focus on the feasibility of detecting observational signatures that are unique to thermonuclear detonation of massive helium shells. Consequently the rate estimated in this work is primarily theoretical and based on binary evolution calculations, although motivated by trying to understand the processes that may occur in thermonuclear supernovae that are not powered primarily by nickel and cobalt decay. Consequently, our calculated rate for ELDD-L-like events has some uncertainties.

In our calculation of the expected event rate as a function of distance, we consider only the mass that is presently contained in stars. This results in an underestimate of supernova rates from systems with long-delay times \citep[e.g.][]{panther2019}, as it does not take into account the total mass that formed into stars - most massive stars that formed in the past will have since exploded as supernovae. As the binary population synthesis calculation depends on knowing the total mass that formed into stars, our calculation underestimates the rate. However, since most of the stellar mass is contained in low-mass stars, the approximation we use for this work is reasonable to obtain a lower limit. 

The model used in this work is comprised of a low-mass CO WD core surrounded by a thick helium envelope. Our rate is calculated based on the incidence of interacting binary systems that may end their lives in such configurations. We consider all systems that involve mergers of low mass CO WDs and helium hybrid WDs, and consequently the rate we estimate is somewhat uncertain and may be a factor of a few larger or smaller, depending on the kind of evolutionary tracks that could give rise to these mergers. However, we point out that any thermonuclear explosion of a system with a massive helium shell is expected to synthesise considerable quantities of $^{48}$Cr, and consequently is expected to emit gamma-rays from the decay of daughter nucleus $^{48}$Cr (e.g. see the model of SN2014J proposed by \citealt{Diehl2015_SN2014J}). This is a unique feature of massive helium shell detonations. Some simulations involving higher mass CO WD cores have been carried out \citep[e.g.,][]{pakmor2021}, which yield interesting synthetic observables that may be of interest in the era of large scale surveys by the likes of the Vera C. Rubin Observatory. 

During the accretion process, nuclear burning on the CO WD core may result in the emission of X-rays giving rise to so-called supersoft X-ray sources \citep{Woods16}. These sources are not readily detectable beyond the Large Magellanic Cloud, making a detectable event very unlikely with existing or future technology. For low-mass CO WDs, such as in our scenario, radiation emitted during the accretion process would be predominantly emitted in the far UV, which is mostly attenuated by interstellar dust and gas in the host galaxy.

Further work is required to understand the explosion properties of configurations involving low mass CO WD cores with thick helium envelopes, especially in the case of low mass CO WD cores. Such systems are of general interest and have been proposed to be a significant source of Galactic antimatter \citep{Crocker17}.

\section{Conclusions}
In this paper we discuss the prospects for observing gamma-ray line emission from supernovae predominantly powered by the decay of the intermediate mass element $^{48}$Cr and its daughter nuclei. We outline an optimized observing strategy based on radiative transfer models of edge-lit double-detonation of a CO WD with a He layer. We compute gamma-ray lightcurves for the two dominant emission lines of $^{48}$V and determine that an optimal, idealized observation strategy would involve a $1-1.5\,\mathrm{Ms}$ exposure beginning at $\sim 12\,\mathrm{days}$ post-explosion. We determine the maximum distances at which a SN explosion exhibiting the V lines similar to that of the model could be observed with 3-sigma significance, taking into account the broad linewidth reducing the sensitivity of each instrument. Finally, we estimate a rate for such events based on binary population synthesis calculations.

While we find the rates of such events to be relatively low, we note that the highly random nature of SN events in the local Universe may mean that we could see a nearby event - such as SN2014J, SN2011fe, or even as close as SN1987A - by virtue of chance. Moreover, we note that the link between the explosion model and the progenitor system rate determined from our binary evolution calculation introduces additional uncertainty. Observation of $^{48}$V emission, a radioisotope only synthesised in significant quantities by thermonuclear detonation of massive helium shells is highly model discriminating. Based on our calculations and the success of observations of gamma-ray lines from other nearby thermonuclear supernovae, we conclude that the potential to directly observe the radioactive decay of Cr and V produced in detonations of massive helium shells on top of white dwarf stars is a real possibility with proposed future gamma-ray instruments. 

\section*{Acknowledgements}
The authors with to thank the reviewer for their constructive comments to improve the paper. 
This research was conducted in Canberra, on land for which the Ngunnawal and Ngambri people are the traditional and ongoing custodians, and in Perth, on land for which the Wadjuk Noongar people are the traditional and ongoing custodians.
Analysis of synthetic spectra were performed using the open-source Python library NumPy \citep{numpy}, and all plots are generated using the open-source Matplotlib Python library \citep{matplotlib}.
FHP is supported by the  Australian Research  Council  (ARC)  Centre  of  Excellence for  Gravitational  Wave  Discovery (OzGrav)  under grant CE170100004.
TS is supported by the German Research Society (DFG-Forschungsstipendium SI 2502/1-1 \& 2502/3-1).
IRS and AJR are supported by the Australian Research Council through grant numbers FT160100028 and FT170100243, respectively.
RMC acknowledges funding from Australian Research Council
award DP190101258 shared with Prof.~Mark Krumholz at the ANU.
SAS acknowledges travel support from the Australian National University’s, Research School of Astronomy \& Astrophysics, Distinguished Visitor Program and the ARC Centre for Excellence for All-sky Astrophysics (CAASTRO) for travel to Australia in 2017, which contributed to part of this work.
This research was undertaken with the assistance of resources and services from the National Computational Infrastructure (NCI), which is supported by the Australian Government, through the National Computational Merit Allocation Scheme and the UNSW HPC Resource Allocation Scheme.
We thank Roland Diehl and Sandra Resch for providing limits on Vanadium lines in SN2014J.
%

%%%%%%%%%%%%%%%%%%%%%%%%%%%%%%%%%%%%%%%%%%%%%%%%%%
\section*{Data Availability}
The data and programs used to analyse the data are available via GitHub at \href{https://github.com/fipanther/Vlineprospects}{https://github.com/fipanther/Vlineprospects}. The repository enables the results presented in this work to be replicated and the spectrum of this model (including optical emission) to be explored if the reader desires. For more information on the models themselves, we direct the reader to the orignal work of \cite{Sim2012} where the model and its optical characteristics were presented. Any use of the provided models must cite the original work of \cite{Sim2012}.

%%%%%%%%%%%%%%%%%%%% REFERENCES %%%%%%%%%%%%%%%%%%
\bibliographystyle{mnras}
\bibliography{PantherBib} % if your bibtex file is called example.bib

% Don't change these lines
\bsp	% typesetting comment
\label{lastpage}
\end{document}